\documentclass[12pt,preprint]{aastex}
\usepackage[dvips]{epsfig}
\usepackage[dvips]{color}

\begin{document}
\shorttitle{\sc AN EXTREMELY METAL-POOR STAR IN THE BO\"{O}TES~I DWARF SPHEROIDAL GALAXY}
\shortauthors{NORRIS ET Al.}

\newcommand{\boo} {Bo\"{o}tes~I}
\newcommand{\boos} {Boo$-$1137}
\newcommand{\cd}{CD$-38^{\circ}\,245$}
\newcommand{\bd}{BD$-18^{\circ}\,5550$}
\newcommand{\kms} {\rm km~s$^{-1}$}
\newcommand{\teff} {$T_{\rm eff}$} 
\newcommand{\logg} {log~$g$} 
\newcommand{\loggf} {log~$gf$} 

\title{BOO--1137 -- AN EXTREMELY METAL-POOR STAR IN THE ULTRA--FAINT
  DWARF SPHEROIDAL GALAXY BO\"{O}TES~I\footnote{Observations obtained
  for ESO program P383.B-0038, using VLT-UT2/UVES}}

\author {JOHN E. NORRIS\altaffilmark{1}, DAVID YONG\altaffilmark{1},
GERARD GILMORE\altaffilmark{2}, AND ROSEMARY
F.G. WYSE\altaffilmark{3}}

\altaffiltext{1}{Research School of Astronomy \& Astrophysics, The
Australian National University, Mount Stromlo Observatory, Cotter
Road, Weston, ACT 2611, Australia; email: jen@mso.anu.edu.au}

\altaffiltext{2}{Institute of Astronomy, University of Cambridge,
Madingley Road, Cambridge CB3 0HA, UK}

\altaffiltext{3}{The Johns Hopkins University, Department of Physics
\& Astronomy, 3900 N.~Charles Street, Baltimore, MD 21218, USA}

\begin{abstract}

We present high-resolution (R $\sim$ 40000), high-$S/N$
(20--90) spectra of an extremely metal-poor giant star {\boos} in the
``ultra-faint'' dwarf spheroidal galaxy (dSph) {\boo}, absolute
magnitude M$_V \sim -6.3$.  We derive an iron abundance of [Fe/H] =
--3.7, making this the most metal-poor star as yet identified in an
ultra-faint dSph. Our derived effective temperature and gravity are
consistent with its identification as a red giant in \boo.

Abundances for a further 15 elements have also been
determined. Comparison of the relative abundances, [X/Fe], with those
of the extremely metal-poor red giants of the Galactic halo shows that
{\boos} is ``normal'' with respect to C and N, the odd-Z elements Na
and Al, the iron-peak elements, and the neutron-capture elements Sr
and Ba, in comparison with the bulk of the Milky Way halo population
having [Fe/H] $\la$ --3.0.  The $\alpha$-elements Mg, Si, Ca, and Ti
are all higher by $\Delta$[X/Fe]~$\sim$~0.2 than the average halo
values.  Monte-Carlo analysis indicates that $\Delta$[$\alpha$/Fe]
values this large are expected with a probability $\sim$ 0.02. The
elemental abundance pattern in Boo--1137 suggests inhomogeneous
chemical evolution, consistent with the wide internal spread in iron
abundances we previously reported.  The similarity of most of the
{\boos} relative abundances with respect to halo values, and the fact
that the $\alpha$-elements are all offset by a similar small amount
from the halo averages, points to the same underlying galaxy-scale
stellar initial mass function, but that {\boos} likely originated in a
star-forming region where the abundances reflect either poor mixing of
supernova ejecta, or poor sampling of the supernova progenitor mass
range, or both.

\end{abstract}


\section{INTRODUCTION}

The discovery and analysis of extremely metal-poor stars (those with
[Fe/H]\footnote{[Fe/H] = log(N(Fe)/N(H))$_{\rm{star}}$ --
log(N(Fe)/N(H))$_{\odot}$} ~$<$~--3.0) in extremely low
luminosity dwarf spheroidal galaxies is changing our perspective on
the early chemical enrichment within these objects, the formation of
the outer regions of the Milky Way halo, and the role of dwarf
spheroidal galaxies (dSph) within the $\Lambda$CDM paradigm of the
manner in which the Milky Way formed.

Initial observations of the brighter dSph (Helmi et al.\ 2006, and
references therein) led to the conclusion that dSph contained no stars
with [Fe/H] $<$ --3.0, while detailed studies of relative abundances
(in particular [$\alpha$/Fe]) showed patterns that were unlike those
of Galactic halo stars in the solar neighborhood (Venn et al.\ 2004,
and references therein; see also Tolstoy, Hill, \& Tosi 2009).  The
paucity of brighter Milky Way dSph and their chemical relative
abundances seem at odds with the CDM paradigm (Klypin et al.\ 1999;
Moore et al.\ 1999) and the concept that these systems are the
building blocks of at least part of the Galaxy's halo. The recent
identification of several ``ultra-faint'' systems, with luminosities
many orders of magnitude fainter than those of the classical dSph,
through analyses of the imaging data from the Sloan Digital Sky Survey
(e.g.~Belokurov et al.\ 2006, 2007) has revitalized discussions of
satellite luminosity and mass functions.  Further, these ultra-faint
systems contain extremely low-metallicity stars (Kirby et al.\ 2008;
Norris et al.\ 2008), perhaps redressing the apparent deficit in
brighter dSph.  Arguably even more interesting than their relevance as
building blocks is to understand the galaxies themselves. Are they the
first objects?  Did they cause reionization?  What do they tell us of
the first stars?  What was the stellar Initial Mass Function (IMF) at
near zero metallicity?  How are the faintest dSph related to more
luminous dSph and the Milky Way?

In a study of several ultra-faint dSph, Kirby et al. (2008) first
reported stars with [Fe/H] $ < -3.0 $ (with metallicities as low as
[Fe/H] $ = -3.3$), based on moderate-resolution spectra in the range
8300--8500\,{\AA}, while Norris et al. (2008) from multi-object
spectroscopy of the {\boo} dSph found a similar result, with the most
metal-poor star having [Fe/H] $= -3.4$ based on the CaII K line
(3933\,{\AA}) in moderate resolution blue spectra.  More recently,
Frebel et al.\ (2009) have obtained the first detailed elemental
abundances, based on high-resolution, high $S/N$ spectra, of extremely
metal-poor stars in the ultra-faint systems, with observations of Com
Ber and U Ma II (M$_V \sim -4, \, -5.5$ respectively).  Of the six
stars studied, one has [Fe/H] $= -3.2$, and another $-3.1$.  Perhaps
the most interesting result of their investigation is that in these
stars [$\alpha$/Fe] is similar to that found in the bulk of Galactic
halo stars.  That is to say, these low metallicity stars in these
faint satellite galaxies were apparently enriched by SNe from a
similar stellar IMF to that which enriched their Milky Way
counterparts.  The most metal-poor -- and possibly oldest -- stars in
the more luminous dSph also show the same level of enhancement of the
$\alpha$-elements (e.g. Koch et al.\ 2008). These results stand in
some contrast to results at higher abundance, in which regime the dSph
member stars have significantly lower levels of [$\alpha$/Fe] than
seen in field halo stars of the same iron abundance (e.g.~Venn et
al.~2004), plausibly reflecting the more extended star formation and
self-enrichment in the dSph in contrast to the field halo (Unavane,
Wyse \& Gilmore 1996).  The ultra-faint systems have sufficiently low
luminosities and low metallicities that one might observe the
anticipated signatures of enrichment by single Type II supernovae,
perhaps even by Population~III massive stars.

The {\boo} system was discovered by Belokurov et al.~(2006) and is an
ultra-faint dSph (M$_V \sim -6.3$, luminosity $\sim 3 \times
10^4$~L$_\odot$; Martin et al.~2008) at a distance of $\sim
65$~kpc. The observed color-magnitude diagram is consistent with an
old, metal-poor system.  We identified extremely metal-poor member
stars in {\boo}, together with a large internal dispersion in
metallicity, through intermediate-resolution spectroscopy of the Ca II
H and K lines (Norris et al.~2008).  We here report follow-up,
high-resolution, high-S/N, observations of the particularly
interesting star {\boos} we identified which lies at $\sim
2$~half-light radii from the center of {\boo}, and for which we had
derived [Fe/H] = --3.4.  In \S2 we present data obtained for this star
using the VLT/UVES system with resolving power R = 40000 and $S/N$ =
20--90.  In \S3 we report chemical abundances from a model atmosphere
analysis of this material for some 16 elements.  {\boos} has [Fe/H] =
--3.7, and relative abundances ([X/Fe]) that are very similar to those
of Galactic halo stars of the same [Fe/H].  We discuss the
implications of our results in \S4.
 
\section{OBSERVATIONS AND REDUCTION}

\subsection{\bf{$ugriz$} Photometry}

{\boos} lies at $\alpha (2000) = 13$\,h $58$\,m $33.8$\,s and $\delta
(2000) = +14^{\circ}\;21'\;08''$.  $ugriz$ photometry is available
from Data Release 7 of the Sloan Digital Sky Survey (Abazajian et al.\
2009, \texttt{http://cas.sdss.org/astrodr7/en/tools/search/}).
Following Belokurov et al.\ (2006), we adopt E(B--V) = 0.02 (and hence
E(g--r) = 0.021 and E(r--z) = 0.027 (see Schlegel, Finkbeiner, \&
Davis 1998)), to obtain (g--r)$_{0}$ = 0.718 $\pm$ 0.009 and
(r--z)$_{0}$ = 0.518 $\pm$ 0.011 for {\boos}.  We also use the
transformation (B--V)$_{0}$ = 1.197$\times$(g--r)$_{0}$ + 0.049,
appropriate for metal-poor red giants (Norris et al.\ 2008), to obtain
(B--V)$_{0}$ = 0.91 $\pm$ 0.01.  These are used in the abundance
analysis in \S3.1.

\subsection{High-resolution Spectroscopy}

{\boos} was observed in Service Mode at the Very Large Telescope (VLT)
Unit Telescope 2 (UT2) with the Ultraviolet-Visual Echelle
Spectrograph (UVES) (Dekker et al.\ 2000,
\texttt{http://www.eso.org/sci/facilities/paranal/instruments/uves/)} during the nights of 2009 April 24 and 25. Ten
individual exposures with an integration time of 46\,min each were
obtained. UVES was used in dichroic mode with the BLUE390 and RED564
settings, covering the wavelength ranges 3300--4520\,{\AA} in the
blue-arm spectra, and 4620--5600\,{\AA} and 5680--6650\,{\AA} in the
lower- and upper- red-arm spectra, respectively. A 1~{\arcsec} wide
slit was used for all observations.

The ten pipeline-reduced spectra were co-added to produce the final
results.  The co-added blue-arm spectrum has a maximum $S/N$ per
$0.027$\,{\AA} pixel $\sim$ $40$ at $4500$\,{\AA}, decreasing to $S/N
\sim 30$ at $4100$\,{\AA} and $S/N \sim 20$ at $3700$\,{\AA}.  Below
$3700$\,{\AA}, the spectrum is only of limited usefulness.  In the
lower red-arm spectrum, the $S/N$ per $0.028$\,{\AA} pixel was
$\sim$~60 throughout, while for the upper red-arm spectrum the $S/N$
per $0.033$\,{\AA} pixel increases from 70 to 90.

\begin{figure}[htbp]
\vspace{1cm}
\begin{center}
\includegraphics[width=10.0cm,angle=-90]{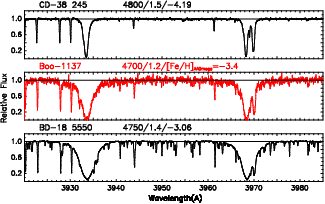}

  \caption{\label{Fig:Spectra} Comparison of the spectrum of {\boos}
  in the region of the Ca II H and K lines with those of the
  metal-poor giants {\cd} ([Fe/H] = --4.2) and {\bd} ([Fe/H] = --3.1).
  The atmospheric parameters {\teff}/log~$g$/[Fe/H] (from Cayrel et
  al.\ (2004) and \S3 of the present work (except for [Fe/H], which comes
  from Norris et al.\ 2008)) are also presented above each
  spectrum.  Note that {\boos} has line strengths, and thus
  abundances, intermediate between those of the two comparison
  objects.  }

\end{center}
\end{figure}

An example of the continuum normalized spectrum of {\boos} in the
region of the Ca~II~K line is shown in Figure~\ref{Fig:Spectra}, where
it is compared with those of the extremely metal-poor giants {\cd} and
{\bd} which have effective temperatures and gravities similar to those
of {\boos}, and abundances [Fe/H] = --4.2 and --3.1, respectively.
Inspection of the figure suggests that {\boos} does indeed have an
abundance consistent with our initial estimate of [Fe/H] = --3.4.

\subsection{Line Strength Measurements}

Beginning with the line list of Cayrel et al.\ (2004), supplemented by
four (non-Fe I) lines that appear unblended in the Sun and Arcturus
and have {\loggf} values in the Vienna Atomic Line Database
(VALD)\footnote{\texttt{http://www.astro.uu.se/$\sim$vald/}}(Kupka et
al.\ 1999), we have used the VLT spectra to measure equivalent widths
of 14 elements in {\boos} in the wavelength range 3800--6650\,{\AA}.
(Two further elements, C and N, are discussed below.)  A comparison of
independent line strength determinations by the first two authors
obtained using techniques described by Norris et al.\ (2001) and Yong
et al.\ (2008) \color{black} is shown in Figure~\ref{Fig:Eqwid}a,b,
where the agreement is quite satisfactory, with an RMS scatter between
the two estimates of 4.0 m{\AA}.  While a small departure from the
one-to-one line is evident in the figure, representing a systematic
difference of a few m{\AA}, we have chosen to average the data, and
present in column (5) of Table 1 line strengths for 152 unblended
lines suitable for model atmosphere abundance analysis.  Lower
excitation potentials, $\chi$, and {\loggf} values are presented in
columns (3) and (4) of the table, taken preferentially from Table 3 of
Cayrel et al.\ (2004), and supplemented by material from the VALD
database.

For heuristic purposes we also show, in Figure~\ref{Fig:Eqwid}c, the
line strengths of {\cd} ([Fe/H] = --4.2) and {\bd} ([Fe/H] = --3.1)
versus those of {\boos}. As one might expect from the comparisons
presented in Figure~\ref{Fig:Spectra}, and recalling that the three
objects have similar effective temperatures and gravities, {\boos} has
line strengths intermediate between those of {\cd} and {\bd}.

\subsection{Radial Velocity}

Radial velocities for {\boos} were measured (over the wavelength range
5160--5190\,{\AA}) by Fourier cross-correlation (using routines in the
FIGARO reduction package\\ (see \texttt{http://www.aao.gov.au/figaro)}
of each of its ten pipeline reduced spectra against a synthetic
spectrum having {\teff} = 4700K, {\logg} = 1.5, [M/H] = --3.5, and
microturbulent velocity $\xi_t$ = 2~{\kms} (computed with the code
described by Cottrell and Norris 1978, and atomic line wavelengths
from VALD). The resulting heliocentric velocity is V$_{\rm r}$ = 99.1
$\pm$ 0.1~{\kms}, with the individual velocities covering the range
98.8--99.5~{\kms}.  The quoted error is the standard error on the
mean, and refers to the internal error of measurement, and does not
include any consideration of the external error, given that spectra of
velocity standards were not obtained as part of this program.
Lucatello et al.\ (2005) find, from a careful study of the metal-poor
subgiant HD~140283, that the external error for UVES is 0.3~{\kms},
while Napiwotzki et al., in a private communication to Norris et
al. (2007), report a value of 0.7~{\kms}.  For the purposes of the
present investigation we shall adopt an external error of 0.5~{\kms},
the mean of these two estimates.

When internal and external errors are taken together, we thus have
V$_{\rm r}$ = 99.1 $\pm$ 0.5~{\kms} for {\boos}, which is consistent
with radial velocity membership of {\boo}, for which Martin et al.\
(2007) report a systemic velocity of 95.6 $\pm$ 3.4~{\kms} and
dispersion of 6.6 $\pm$ 2.3~{\kms}. It also agrees with the previous
value reported for this star by Norris et al. (2008), when their cited
value is placed on the system of Martin et al.\ (by requiring that the
mean velocities of the two investigations agree), which then becomes
104 $\pm$ 7~{\kms}.

\section{ABUNDANCE ANALYSIS}

\subsection {Effective Temperature and Surface Gravity}

{\teff} and {\logg} were determined from (g--r)$_{0}$ and (r--z)$_{0}$
by assuming that {\boos} lies on the red giant branch and iteratively
using the synthetic $ugriz$ colors of Castelli\\
\texttt{(http://wwwuser.oat.ts.astro.it/castelli/colors/sloan.html)}
and the Yale--Yonsei Isochrones (Demarque at al.\ 2004,
\texttt{http://www.astro.yale.edu/demarque/yyiso.html}), with an age
of 12 Gyr. Similarly, the (B--V)$_{0}$ value for {\boos} was used
together with these isochrones to provide another estimate of the
parameters.  The process required a knowledge of the chemical
abundance.  In practice, first estimates of {\teff} and {\logg}, based
on the initial value of [Fe/H] = --3.4 from Norris et al.\ (2008),
were adopted in the model atmosphere abundance analysis described
below, and an iterative procedure followed using the new value of
[Fe/H], until convergence was obtained.  This happened after one
iteration.  The individual values of {\teff} and {\logg} lie in the
ranges 4640--4760K and 1.0--1.3 dex, respectively.  Our final adopted
values are {\teff} = 4700K and {\logg} = 1.2.  It is difficult to
estimate the systematic errors in these parameters: in what follows we
shall somewhat arbitrarily adopt $\Delta${\teff} = 200K and
$\Delta${\logg} = 0.3.

\subsection {Relative Abundances from Atomic Features}

We determined the abundance of key elements beginning with iron -- the
canonical measure of metallicity.  Model atmospheres were taken from
the NEWODF grid of ATLAS9 models (plane-parallel, one-dimensional
(1D), local thermodynamic equilibrium (LTE)) of Castelli \& Kurucz
(2003, \texttt{http://wwwuser.oat.ts.astro.it/castelli/grids.html}).
The particular grid of models used was $\alpha$-enhanced,
[$\alpha$/Fe] = +0.4, and computed assuming a microturbulent velocity
of $\xi_t$ = 2~{\kms}. Interpolation within the grid was performed
when necessary to produce models with the required {\teff}, {\logg},
and [M/H]. The interpolation software, kindly provided by Dr Carlos
Allende Prieto, has been used extensively (e.g., Reddy et al.\ 2003
and Allende Prieto et al.\ 2004).

The model atmospheres were used in conjunction with two versions of
the LTE stellar line analysis program MOOG (Sneden 1973).  The first
was the 2009 standard version then available at
\texttt{http://verdi.as.utexas.edu/moog.html}, while the second is
currently under development, and uses a more rigorous treatment of
continuum scattering.  The latter version was generously made
available to us by Prof. Chris Sneden.  We tested both versions by
analyzing the equivalent widths of the 35 metal-poor stars of Cayrel
et al.\ (2004), for lines having $\lambda \ga$ 3750\ {\AA}.  Adopting
their atmospheric parameters, we found small but significant
convergence in our abundances compared with theirs, when using the
more recent version.  That is to say, the agreement between our and
their abundances went from very good to excellent.  In what follows,
therefore, we shall present results obtained by using the newer
version of the code.  This will prove useful in \S4.2, where we shall
compare our abundances for {\boos} with those for the Galactic halo
giants of Cayrel et al.\ (2004).  We shall comment below on the
abundance differences that resulted, for the set of lines that we
analyzed, between the two versions.

The microturbulent velocity, $\xi_t$, was determined in the usual way
by requiring the abundances from Fe\,{\sc i} lines to be independent
of their reduced equivalent width, log($W_{\lambda}$/$\lambda$). For
{\boos} we obtain $\xi_t$ = 2.2 $\pm$ 0.1~{\kms}.  Determination of
the Fe abundance offers a check on the adopted stellar parameters and
the assumptions underlying the model atmospheres and line analysis,
principally via the ionization and excitation balance.  Concerning
ionization equilibrium, the mean Fe abundances derived from neutral
lines and from ionized lines provide a check on the adopted surface
gravity.  Additionally, differences between the abundances from
Fe\,{\sc i} and Fe\,{\sc ii} lines may represent departures from LTE
(e.g. Thev\'{e}nin \& Idiart 1999; but see also Gratton et
al. 1999). For {\boos}, these abundances agree within 0.06 dex, i.e.,
ionization equilibrium is satisfied, and this suggests our surface
gravity is appropriate and departures from LTE are small.  

Regarding excitation equilibrium, we note that other studies of
metal-poor giants have reported a systematic trend between the
abundance from Fe\,{\sc i} lines and lower excitation potential,
$\chi$, (e.g., Cayrel et al.\ 2004, Lai et al.\ 2008, and Cohen et
al.\ 2008).  Cayrel et al.\ (2004) and Lai et al.\ (2008) found that
the trend between abundances from Fe\,{\sc i} lines and lower
excitation potential is alleviated, or indeed removed, by excluding
lines with $\chi <$ 1.2~eV.  In our analysis, we also find a trend
between the abundances from Fe\,{\sc i} lines and lower excitation
potential when considering all lines -- with slope $-$0.12 dex/eV,
which is compatible in both sign and magnitude with the (lower) values
reported by Lai et al.\ (2008).  When considering only lines with
$\chi >$ 1.2~eV, however, there is no statistically significant trend
(even at the 1$\sigma$ level), as reported in previous
studies\footnote{This has clear implications for the use
of the excitation equilibrium, as determined via 1D, LTE analysis, as a
means of determining {\teff} for metal-poor red giants in terms of
``excitation temperature''.  In our initial efforts to do this, we
were forced to significantly lower, and arguably
non-physical/implausible, {\teff} in order to remove the dependence of
iron abundance on lower excitation potential when we used lines at all
values of $\chi$, as opposed to those obtained when we removed lines
having lower values.}.

Having performed these validity checks and obtained a measure of the
microturbulent velocity, we then computed abundances for atomic
features with measured equivalent widths using the adopted model
atmosphere and MOOG. Some further details should, however, be
noted. First, lines of Sc\,{\sc ii}, Mn\,{\sc i}, and Co\,{\sc i} are
affected by hyperfine splitting (HFS). In our abundance analysis, HFS
was treated appropriately using the parameters from Kurucz \& Bell
(1995).  In the case of Mn, Cayrel et al.\ (2004) also noted that the
resonance triplet $a^{6}{\rm S}-z^{6}{\rm P}^{0}$ at 403\,nm yields
abundances systematically too low by 0.4 dex compared with results
from other Mn lines. We have thus increased the Mn abundances in
Tables 1 and 2, and throughout this work, by that amount.  Finally,
for the elements O, Zn and Eu, which are not measurable in our
spectra, but which are observed at higher metallicity and have
particular significance in comparison with models of galactic chemical
enrichment, we computed abundance limits based on the O\,{\sc
i}~6300.30\,{\AA}, Zn\,{\sc i}~4810.53\,{\AA}, and Eu\,{\sc
ii}~4129.72\,{\AA} lines, respectively, by adopting upper limit
equivalent widths of 10~m\,{\AA}.

Our abundances are presented in Table 2, where columns (1)--(5)
contain the species, the number of lines measured (or, alternatively,
that synthetic spectra were compared with observations),
log($\epsilon$(X))\footnote{log~$\epsilon$(X) =
log(N(X)/N(H))$_{\rm{star}}$ + 12.00}, its error, and relative
abundance [X/Fe], respectively.  (In order to compute the relative
abundances we adopted the solar abundance data of Asplund et al.\
(2005)). While we present abundances obtained using the revised
version of MOOG, discussed above, in Table 2, we note that for our set
of equivalent widths (with $\lambda \ga$ 3750\ {\AA}), the revised
version produces a value of [Fe/H] lower than the standard version by
0.03 dex, and relative abundances lower on average by 0.02 dex for the
16 species studied.  As expected, the difference increases as one goes
to shorter wavelength.  For the atomic lines discussed here, the
largest difference was --0.09 dex, for Al I, for which we observed
only two lines -- at 3944.00{\AA} and 3968.52{\AA}.

It will be important in the discussion of relative abundances in \S4.2
to appreciate the systematic differences between the abundances
presented here and those of Cayrel et al.\ (2004).  In order to do
this, we analyzed the equivalent width data in Table 3 of Cayrel et
al.\, adopting their atmospheric parameters, and using our techniques.
For stars having [Fe/H] $<$ --3.0, the comparison of the
log~$\epsilon$ values for the atomic species in Table 8 of Cayrel et
al.\ (2004) with our results (in the sense [Cayrel et al. -- present
work]) produces the following mean abundance differences:
$\Delta$[Fe/H] = 0.023 $\pm$ 0.003, $\Delta$[Na I/Fe] = --0.071 $\pm$
0.040, $\Delta$[Mg I/Fe] = --0.049 $\pm$ 0.007, $\Delta$[Al I/Fe] =
--0.030 $\pm$ 0.009, $\Delta$[Ca I/Fe] = --0.008 $\pm$ 0.006,
$\Delta$[Sc II/Fe] = --0.012 $\pm$ 0.009, $\Delta$[Ti I/Fe] = 0.033
$\pm$ 0.004, $\Delta$[Ti II/Fe] = --0.019 $\pm$ 0.006, $\Delta$[Cr
I/Fe] = --0.001 $\pm$ 0.003, $\Delta$Fe II/Fe] = --0.027 $\pm$ 0.007,
$\Delta$[Co I/Fe] = 0.005 $\pm$ 0.002, $\Delta$[Ni/Fe] = 0.016 $\pm$
0.0.003.  We regard this agreement, which is independent of adopted
solar abundances, as very satisfactory.

We conclude this section by noting that abundances of both neutral and
singly ionized lines of Ti are presented in the table, and permit a
further check of the surface gravity and the presence of departures
from LTE. One finds that abundances for the two ionization states
agree within 0.08 dex, which, given their relative uncertainties, we
regard as agreement (see section \S3.4 on error analysis).

\subsection {Relative Abundances from Molecular Features}

Our spectra for {\boos} include features of CH and NH, which permit us
to determine the abundances of carbon and nitrogen.  For carbon, we
compared observed and synthetic spectra (generated with MOOG) of the
(0,0) and (1,1) bands of the $A-X$ electronic transition of the CH
molecule in the interval 4250--4330\,{\AA}, while for nitrogen we used
the (0,0) and (1,1) bands of the $A-X$ electronic transition of the NH
molecule in the range 3340--3400\,{\AA}.  (As noted above in \S2.2,
the $S/N$ is relatively poor in the latter wavelength range, which is
reflected in the lower accuracy of our deduced nitrogen abundance.)

For carbon, we used the Plez et al.\ (2008) line list, and a
dissociation energy of 3.465~eV.  The abundance of C is weakly
dependent on the assumed O abundance: since we have only an upper
limit to the latter, the synthetic spectra were computed using the
upper limit in Table 2, ([O/Fe] $< 1.94$) and also with [O/Fe] = +0.5,
consistent with the values observed in Galactic halo stars.  For this
range of O abundance, the inferred C abundance does not change.  We
adjusted the abundance of C until the synthetic spectrum matched the
observed one, as may be seen in Figure~\ref{Fig:Carbon}.  We find
[C/Fe] = 0.25 $\pm$ 0.2 for {\boos}.

For nitrogen, the line list was taken from Johnson et al.\ (2007), in
which the Kurucz-$gf$ values were reduced by a factor of 2.  We also
adopted the dissociation potential of 3.450~eV.  Given the poorer
$S/N$ noted above, the observed spectrum was smoothed with a 5-pixel
boxcar function to increase the signal-to-noise ratio.
Figure~\ref{Fig:Nitrogen} shows the comparison between observed and
synthetic spectra, from which we infer [N/Fe] = 1.1 $\pm$
0.3.  That is, {\boos} possesses a significantly enhanced relative
nitrogen abundance compared with the solar value.

We noted above that the newer version of the spectrum analysis code
MOOG produces more reliable abundances at shorter wavelengths than
those from the older one, and that for the atomic features analyzed
above (with $\lambda \ga$ 3750\ {\AA}) the largest difference was 0.09
dex. At the wavelength of the NH features, 3340--3400\,{\AA}, the
difference is substantially larger, with the newer version producing an
abundance that is lower by $\Delta$[N/Fe] = 0.4.

\subsection {Abundance Errors}

The determination of the abundances in Table 2 is also subject to
uncertainties in the adopted atmospheric parameters.  We have
estimated these errors by repeating the abundance analysis and varying
the parameters, one at a time, by $\Delta${\teff} = +200K,
$\Delta${\logg} = +0.3, $\Delta$[M/H] = +0.3, and $\Delta\xi_t$ =
+0.2~{\kms}.  The results are presented in Table 3, where columns
(2)--(5) contain the individual errors, and the final row shows the
accumulated error when the four uncertainties are added quadratically.
To obtain total error estimates, which we shall use in \S4.2, we
proceed as follows. Noting that some of the errors in Table 2
involving small numbers of lines are implausibly small, we replace the
value in Table 2 (s.e.$_{\log\epsilon}$) by max(s.e.$_{\log\epsilon}$,
0.20/$\sqrt{N_{\rm lines}}$), where the second term is what one might
expect from a set of N$_{\rm lines}$ having dispersion 0.20 dex (the
value we obtained for the abundance dispersion of our Fe I lines).  We
then quadratically add the updated random error and the systematic
error in Table 3 to obtain the final total error.

\section{DISCUSSION}

{\color{black} \subsection {{\boo} Membership}

{\boos} lies 24{\arcmin} from the center of {\boo}, corresponding to
1.9 half-light radii (following Martin et al.\ 2008).  Its radial
velocity (V$_{\rm r}$ = 99.1~{\kms}) lies within 4~{\kms} of the
systemic velocity of the system (\S2.3), while its ionization balances
of both {\ion{Fe}{2}}/{\ion{Fe}{1}} and {\ion{Ti}{2}}/{\ion{Ti}{1}}
are consistent with its being a red giant with [Fe/H] = --3.7 (\S3.2).
We refer the reader to the discussion by Norris et al.\ (2008,
Footnote 12), based on the discovery statistics of the HK and HES
metal-poor star surveys (Beers et al.\ 1992; Christlieb et al.\ 2008),
of the likelihood that such an extremely metal-poor giant belonging
to the Galactic halo would lie in the direction of {\boo}: they
concluded that 0.02 such stars might be expected.  All of these
facts confirm to us that {\boos} is a member of the system.

\subsection {Relative Abundances}

Figure~\ref{Fig:XFe} presents [X/Fe] as a function of [Fe/H], in the
range --4.5 $<$ [Fe/H] $<$ --2.55, for 12 representative elements in
{\boos} (the open red circle) and some $\sim$~30 metal-poor Galactic
red giants having high-resolution, high $S/N$, abundance analyses,
together with data for dSph systems known to have member stars in the
range [Fe/H] $\la$ --3.1. For the halo stars, for reasons of
homogeneity, we restrict the data to the results of the First Stars
consortium (Cayrel et al.\ 2004; Spite et al.\ 2005; and Fran{\c c}ois
et al.\ 2007), while for the ultra-faint dSph we plot the results of
Frebel et al.\ (2009; Com Ber and U Ma II, filled red circles) and for
the more luminous Sextans dSph the data of Aoki et al.\ (2009) (filled red
triangles)\footnote{We have modified the literature values to correct
for differences in adopted solar abundances between them and the
present work.  On the scale of Figure~\ref{Fig:XFe}, however, this
effect is small: for example, if one considers the abundances of
Cayrel et al.\ (2004) and Spite et al.\ (2005) the mean difference in
relative abundances caused by difference in the adopted solar values,
over the elements in Table~2, is --0.02 dex.  The maximum absolute
difference, for Na, is 0.11 dex.}.

It is important to recall that all of the data in the figure have been
determined using 1D model atmospheres and the LTE approximation.  This
should be borne in mind when comparison is made with predictions of
stellar evolution and galactic chemical enrichment models.  For an
appreciation of modifications that need to be made to the present
abundances to take into account the role of more realistic 3D models
and non-LTE effects we refer the reader to Asplund (2005), and
references therein. That said, the question we are interested to
address here is the similarity or otherwise between the most
metal-poor dSph stars and those of the Galactic halo.  Insofar as we
have established that our 1D, LTE techniques reproduce the abundances
of Cayrel et al.\ (see \S3), Figure~\ref{Fig:XFe} suffices for our
needs.

As one moves from top to bottom in Figure~\ref{Fig:XFe}, six pairs of
related elements are plotted -- representing the CNO group, the light
odd-Z elements, the $\alpha$-elements, the Fe-peak below (Cr and Mn)
and above (Co and Ni) iron, and the neutron-capture elements.  Initial
inspection of the figure suggests an overall similarity between the
relative abundances of {\boos} and those of the Galactic halo at
[Fe/H] $\sim$ --3.5.  We make the following points: (1) For [C/Fe] and
[N/Fe], given the large dispersion in the measured field star
abundance ratios, and our large observational errors, the results for
Boo-1137 are consistent with those for the Galactic halo ; (2) For
[Cr/Fe], [Mn/Fe], [Co/Fe], and [Ni/Fe] (the Fe-peak elements for which
we have data) the 1$\sigma$ error bars all overlap the Cayrel et al.\
(2004) regression lines (their Table 9), supporting the view that
similar processes and enrichment occurred for the material from which
{\boos} and the Galactic halo formed. It also suggests that the
present techniques produce results on the same system as those of
Cayrel et al.\ (2004), strengthening the similar conclusion reached in
\S3.2, based on our analysis of the Cayrel et al.~data; (3) For
the light odd-Z elements, the {\boos} [Na/Fe] and [Al/Fe] data overlap
those of the halo at the 1$\sigma$ level; (4) Of the two heavy
neutron-capture elements, [Sr/Fe] lies within the values for halo
stars, while [Ba/Fe] appears high. Given the complicated trend and
scatter in [Ba/Fe] values seen in the figure, and the paucity of stars
below [Fe/H] = --3.5, more data would be required to address this
issue; and (5) For the representative $\alpha$-elements, the [Mg/Fe]
and [Ca/Fe] values of {\boos} are both higher, by approximately twice
their errors of measurement, than the Cayrel et al.\ (2004) regression
lines of the Galactic halo.

Given that we also have abundances for the $\alpha$-elements Si and
Ti, we examine the final point more closely in
Figure~\ref{Fig:Alphas}, which shows results for [Mg/Fe], [Si/Fe],
[Ca/Fe], and [Ti/Fe], together with [$\alpha$/Fe] (the average of
[Mg/Fe], [Ca/Fe], and [Ti/Fe])\footnote{Here we adopt [Ti/Fe] = ([Ti
I/Fe] + [Ti II/Fe])/2, and exclude [Si/Fe] from the average because it
is generally based on only one, relatively strong, line.  For each of
the other elements five or more lines are available.} as a function of
[Fe/H]. Also shown in Figure~\ref{Fig:Alphas} are the regression lines
of Cayrel et al.\ (2004, their Table 9), supplemented by our linear
least squares fits for Ti I, Ti II, and [$\alpha$/Fe] (not given
individually by Cayrel et al.).  There are interesting similarities in
the panels of Figure~\ref{Fig:Alphas}: the most obvious (and relevant
for the present discussion) is that all of the {\boos} relative
abundances are larger that those of the Galactic halo at the [Fe/H]
value of {\boos}.  Is this significant?  To address this problem we
proceed as follows.  Rows (1)--(6), columns (1)--(7), of Table 4
present the relevant input data: columns (1)--(3) contain the atomic
species involved, the RMS scatter for material having [Fe/H] $<$ --3.0
about the regression lines in Figure~6 (from Cayrel et al.\ (2004) and
the present work), and the resulting values of the relative abundances
of the Galactic halo at [Fe/H]$_{\rm Boo-1137}$ = --3.66.  Columns
(4)--(7) show for {\boos} its relative abundance and error, the
distance it falls above the halo line, and that distance expressed in
units of the star's abundance error.  We then use Monte Carlo analysis
to address the following question: if one draws putative stars at
random from a gaussian distribution for a Galactic halo having the
abundance dispersions in column (2), and superimposes on that a random
gaussian error corresponding to the observational errors of {\boos} in
column (5), what fraction of the resulting abundances would be at
least as large as the observed distance of {\boos} above the halo
lines (presented in column (6)).  The results are presented in the
final column of Table 4.  One sees in rows (1)--(6) that, taken
individually, the probabilities of {\boos} lying above the Galactic
halo values are in the range 0.03--0.14, broadly consistent with the
results in column (7) of the table. The most stringent condition, as
might be expected, comes from the average of the $\alpha$-elements in
the sixth row: the likelihood of finding the observed enhancement of
the averaged $\alpha$-elements is 0.017.

One might also ask the question : what is the probability of {\it all}
of the {\boos} [Mg/Fe], [Si/Fe], [Ca/Fe], [Ti I/Fe], and [Ti II/Fe]
values falling above their respective Galactic halo lines.  The
answer, based on the Monte Carlo simulations, is that the fraction of
relative abundances as large as seen for the five $\alpha$ species is
3~$\times$~10$^{-6}$. This assumes, however, that the abundances of
each of the above species is independent of all of the others, which
is not the case, given the observed propensity of several of them to
often have correlated behaviour.  An example of such correlation is
highlighted in Figure~\ref{Fig:Alphas} where the red star represents
the data of Cayrel et al.\ (2004) for CS22968--014, which for all
species fall below the least-squares lines of best fit to the Cayrel
et al. data.  Had we performed the same test for CS22968--014 we would
have concluded that the likelihood of all of the five species lying so
far below the regression lines is 2~$\times$~10$^{-4}$.  (We note for
completeness that McWilliam et al.\ (1995) also analyzed CS22968--014,
and first reported the low relative abundances of its
$\alpha$-elements.)  The reader will see other examples of this effect
in Figure~\ref{Fig:Alphas}. Given such correlated behavior, we shall
not consider further the test in this paragraph, which is
inappropriate.

The reader may recall from \S3.2 that we found small systematic
differences between the relative abundances of Cayrel et al.\ (2004)
and those we obtained using our techniques.  Those relevant here are
$\Delta$[Mg I/Fe] = --0.049, $\Delta$[Ca I/Fe] = --0.008, $\Delta$[Ti
I/Fe] = 0.033, and $\Delta$[Ti II/Fe] = --0.019, in the sense [Cayrel
et al.\ -- present work].  If we adjust the results in column (4) for
these four species and [$\alpha$/Fe] to take the corrections into
account, the fraction of Monte Carlo simulations having [$\alpha$/Fe]
lying at or above the observed {\boos} values then increases to 0.024. 

The above considerations depend on a knowledge of the dispersions in
relative abundance in the Galactic halo.  In the range --4.1 $<$
[Fe/H] $<$ --3.1, Cayrel et al.\ (2004) report dispersions for
[Mg/Fe], [Si/Fe], [Ca/Fe], and [Ti/Fe] about their regressions against
[Fe/H] of 0.11, 0.20, 0.11, and 0.09 dex, respectively.  Inspection of
Figure~\ref{Fig:Alphas} suggests that given the decreasing sample size
as one moves to lower abundances one should proceed with caution.  For
example, there are only five objects in the figure that have [Fe/H]
$<$ --3.5.  More data are clearly needed before one regard these
estimates as definitive.

With this caveat in mind, we note that these dispersions for [Fe/H]
$<$ --3.1 are somewhat larger than reported for metal-poor dwarfs in
the range --3.0 $\la$ [Fe/H] $\la$ --2.0: by Magain (1989) --
``extremely small (if any)'', by Nissen et al.\ (1994) --
$\sigma$[Mg/Fe] = 0.06 dex, and by Arnone et al.\ (2005) --
$\sigma$[Mg/Fe] = 0.06 dex.  As discussed by these authors, and also
by Argast et al.\ (2002), the dispersion of these elements at lowest
abundance places strong constraints on the yields of SNe, the IMF, and
galactic chemical enrichment at the earliest times.  We shall return
to the implications of this point below.}

We note in concluding this section that Feltzing et al.\ (2009) have
recently reported an anomalously large value of [Mg/Ca] = 0.73 (at
[Fe/H] = --2.0) for one of seven stars they have observed in {\boo}
(all with [Fe/H] $>$ --3.0).  The value we obtain for {\boos} is
[Mg/Ca] = --0.05, which is not too dissimilar from the mean value of
0.11 $\pm$ 0.06 that one obtains for their other six stars.

\subsection{The Evolution and Chemical Enrichment of {\boo}}

What are the implications from Figures~\ref{Fig:XFe} and
\ref{Fig:Alphas} for the manner in which chemical enrichment occurred
in {\boo}?  As noted in \S4.2 above, the initial impression of overall
similarity between the abundances of {\boos} and those of the Galactic
halo in Figure~\ref{Fig:XFe} is perhaps not too surprising.
Intuitively, one might expect that for abundances as low as [Fe/H] =
--3.7, and an old stellar population, it is more likely to find
abundance patterns driven by enrichment from core-collapse supernovae
from massive-star progenitors, without the later modification by Type
SN Ia\footnote{We implicitly ignore here possible abundance
contamination effects that might result from mass transfer in a binary
system as, for example, is believed to have occurred in the CEMP-s
class of metal-poor stars (see Beers \& Christlieb 2005).}.  This of
course will depend on the rate and duration of star formation -- only
the earliest stars will have enrichment from only core-collapse
supernovae, and, should galactic chemical enrichment occur for long
enough, one will see the SN~Ia signature downturn of [$\alpha$/Fe] vs
[Fe/H] to more solar-like values as one moves from lower to higher
[Fe/H].  In the Sculptor and Draco dSph, for example, the onset of the
decrease is evident already at [Fe/H] $\sim$ --2.0, compared with
--1.0 for the Galactic halo (see e.g. Tolstoy et al.\ 2009, their
Figure 11, and Cohen \& Huang 2009, respectively).

The chemical evolution of {\boo} probably involved a relatively
short-lived epoch of star formation and self-enrichment in a dark
matter dominated potential, terminated by catastrophic gas loss in
(Type~II)-supernova-driven winds (Saito 1979; Wyse \& Silk 1985; Dekel
\& Silk 1986) in which the bulk of the initial (gaseous) baryonic mass
was lost. As noted earlier, the color-magnitude diagram of {\boo}
(Belokurov et al.\ 2006) is consistent with an old, metal-poor
population.  Forming stars would then have been chemically enriched by
only core-collapse supernovae, resulting, for example, in enhanced
[O/Fe] and [$\alpha$/Fe] compared with the solar value.  The actual
value of the relative enhancement depends on the mix of masses of the
SNe progenitors, since model SNe yields show that more massive
progenitors produce relatively more intermediate-mass elements than
iron for many elements, so that an IMF biased towards the most massive
stars will, with good sampling of the IMF and good mixing so that an
IMF-average is achieved in star-forming regions, provide higher mean
[O/Fe], [$\alpha$/Fe], etc. (see e.g.~Wyse \& Gilmore 1992; Argast et
al.\ 2002; Kobayashi et al.\ 2006).  The low stellar-mass and low
level of enrichment of {\boo} mean that it would not be surprising if
either the underlying IMF were not well-sampled in star-forming
regions, or that they were not well-mixed, or both.

Our demonstration that the $\alpha$-elements appear enhanced with
respect to iron in Boo-1137, relative to the mean of the metal-poor
halo stars, by $\Delta[\alpha$/Fe] $\sim$ 0.2 is consistent with
enrichment of the material from which {\boos} formed being biased
towards the ejecta of SNe with more massive progenitors than the bulk
of the metal-poor Galactic halo. This could reflect a biased
underlying galaxy-scale IMF, or incomplete mixing and/or poor sampling
of an invariant IMF, or be due to the shorter lifetimes of the most
massive core-collapse progenitors in an invariant IMF.

We also know that {\boo} exhibits a large range in heavy element
abundance -- from samples of 16 and seven members of {\boo} Norris et
al.\ (2008) and Feltzing et al.\ (2009) report ranges of 1.7 and 0.9
dex, respectively.  Assuming a normal underlying galaxy-scale mass
function, for which the stellar mass at $\sim 10$~Gyr after star
formation burst is $\sim 50$\% of the stars formed, and for which
there is one $\sim 10$~M$_\odot$ SN progenitor for every $\sim
100$~M$_\odot$ of stars formed, and one 25~M$_\odot$ SN progenitor for
every $\sim 1000$~M$_\odot$ formed, we can envisage that the early
evolution of {\boo} formed $\sim 10^5~$M$_\odot$ in stars and $\sim
200$~Type II supernovae.  Nissen et al.\ (1994) argued that the small
dispersion, $\sigma$[Mg/Fe] = 0.06, they obtained for the Galactic
halo is consistent with evolution in a well-mixed region with
enrichment from some 25 SNe in the mass range 13--40~M$_{\odot}$ in a
``preceding generation of totally about 2 $\times$ 10$^{4}$ stars''.
{\color{black} (From their Table 8, one would infer that the number of
stars would be smaller by a factor of a few for the somewhat larger
dispersions reported by Cayrel et al.\ (2004) and discussed
above.)}. Given the resultant small number of cells this would imply
if applied to {\boo}, it is not difficult to envisage incomplete
mixing and poor sampling of the IMF across {\boo}, to give the
apparent higher values of [$\alpha$/Fe] in {\boos}.  This is the most
conservative of the three suggested reasons behind the elemental ratio
enhancements we gave above, and we advocate this
conclusion. Individual star-forming regions may well be internally
mixed, while still poorly sampling the core-collapse supernovae IMF.
A prediction of this would be that as more data are collected, one
will observe the signature of a small number of internally well-mixed
cells having preferred abundances.

There remains the objection that these conclusions are based on
results for just one star. Statistical chance in selecting one star
from a possibly diverse parent population, or in enriching one new
star from an inhomogeneous star forming region, may undermine our
conclusions. The analyses of larger samples of extremely metal-poor
stars in {\boo} and in other dSph are awaited with much anticipation.

\acknowledgements

 Studies at RSAA, ANU, of the most metal-poor stellar populations are
 supported by Australian Research Council grants DP0663562 and
 DP0984924, which J.E.N. and D.Y. are pleased to acknowledge.
 R.F.G.W. acknowledges grants from the W.M. Keck Foundation and the
 Gordon \& Betty Moore Foundation, to establish a program of
 data-intensive science at the Johns Hopkins University.  We also
 gratefully thank Carlos Allende Prieto, Bertrand Plez, and Chris
 Sneden for proving us with software that facilitated our
 investigation.\\

\noindent{\it Facilities:} {VLT:Kueyen(UVES)}

\clearpage

\begin{deluxetable}{lllrrr}
\tablecolumns{6}
\tablewidth{0pt}
\tablecaption{\label{Tab:Linelist} AVERAGED EQUIVALENT WIDTHS, UPPER LIMITS, AND LINE-BY-LINE ABUNDANCES OF BOO--1137}
\tablehead{
  \colhead{}    & \colhead{$\lambda$} & \colhead{$\chi$} & \colhead{$\log gf$} &
  \colhead{$W_{\lambda}$} & \colhead{$\log\epsilon$} \\ 
 \colhead{Species} & \colhead{({\AA})}    & \colhead{(eV)}  & \colhead{(dex)}    &
  \colhead{(m{\AA})} & \colhead{(dex)} \\
  \colhead{(1)} & \colhead{(2)}    & \colhead{(3)}  & \colhead{(4)}    &
  \colhead{(5)} & \colhead{(6)} \\
  }
\startdata
Na I   & 5889.951 &   0.00 &   0.11 &   93.9 &   2.42 \\
       & 5895.924 &   0.00 &  $-$0.19 &   77.8 &   2.48 \\
Mg I   & 3829.355 &   2.71 &  $-$0.21 &  115.7 &   4.03 \\
       & 3832.304 &   2.71 &   0.15 &  153.1 &   4.42 \\
       & 3838.290 &   2.72 &   0.41 &  152.1 &   4.15 \\
       & 4351.906 &   4.34 &  $-$0.52 &   44.6 &   4.89 \\
       & 4702.991 &   4.35 &  $-$0.67\tablenotemark{a} &   23.4 &   4.64 \\
       & 5172.684 &   2.71 &  $-$0.38 &  130.2 &   4.17 \\
       & 5183.604 &   2.72 &  $-$0.16 &  142.1 &   4.19 \\
       & 5528.405 &   4.34 &  $-$0.34 &   22.2 &   4.24 \\
Al I   & 3944.006 &   0.00 &  $-$0.64 &   79.4 &   2.18 \\
       & 3961.520 &   0.01 &  $-$0.34 &   86.2 &   2.01 \\
Si I   & 3905.523 &   1.91 &  $-$1.09 &  147.6 &   4.62 \\
Ca I   & 4226.728 &   0.00 &   0.24 &  135.6 &   2.74 \\
       & 4283.011 &   1.89 &  $-$0.22 &   22.8 &   3.17 \\
       & 4454.779 &   1.90 &   0.26 &   52.5 &   3.24 \\
       & 5588.749 &   2.52 &   0.21 &   12.7 &   3.11 \\
       & 6102.723 &   1.88 &  $-$0.79 &    8.4 &   3.14 \\
       & 6122.217 &   1.89 &  $-$0.32 &   24.1 &   3.21 \\
       & 6162.173 &   1.90 &  $-$0.09 &   33.1 &   3.17 \\
       & 6439.075 &   2.52 &   0.47 &   19.2 &   3.03 \\
       & 6493.781 &   2.52 &  $-$0.11\tablenotemark{a} &   22.5 &   3.69 \\
Sc II  & 4246.822 &   0.31 &   0.24 &   74.1 &  $-$0.93 \\
       & 4314.083 &   0.62 &  $-$0.10 &   35.3 &  $-$0.78 \\
       & 4400.389 &   0.61 &  $-$0.54 &   22.9 &  $-$0.59 \\
Ti I   & 3998.636 &   0.05 &  $-$0.06 &   48.5 &   1.88 \\
       & 4981.731 &   0.84 &   0.50 &   27.4 &   1.80 \\
       & 4991.065 &   0.84 &   0.38 &   21.5 &   1.79 \\
       & 4999.503 &   0.83 &   0.25 &   22.7 &   1.93 \\
       & 5210.385 &   0.05 &  $-$0.88 &   12.4 &   1.81 \\
Ti II  & 3759.296 &   0.61 &   0.27 &  141.2 &   1.55 \\
       & 3761.323 &   0.57 &   0.17 &  156.3 &   1.89 \\
       & 3913.468 &   1.12 &  $-$0.41 &  100.6 &   1.78 \\
       & 4012.385 &   0.57 &  $-$1.75 &   73.3 &   1.85 \\
       & 4290.219 &   1.16 &  $-$0.93 &   68.8 &   1.62 \\
       & 4300.049 &   1.18 &  $-$0.49 &   82.2 &   1.44 \\
       & 4394.051 &   1.22 &  $-$1.77 &   29.5 &   1.87 \\
       & 4395.033 &   1.08 &  $-$0.51 &   98.2 &   1.63 \\
       & 4399.772 &   1.24 &  $-$1.22 &   57.3 &   1.81 \\
       & 4418.330 &   1.24 &  $-$1.99 &   23.1 &   1.98 \\
       & 4443.794 &   1.08 &  $-$0.70 &   92.6 &   1.70 \\
       & 4444.558 &   1.12 &  $-$2.21 &   19.0 &   1.95 \\
       & 4450.482 &   1.08 &  $-$1.51 &   56.8 &   1.89 \\
       & 4464.450 &   1.16 &  $-$1.81 &   43.0 &   2.07 \\
       & 4468.507 &   1.13 &  $-$0.60 &   96.8 &   1.74 \\
       & 4501.273 &   1.12 &  $-$0.76 &   80.6 &   1.56 \\
       & 5188.680 &   1.58 &  $-$1.05 &   42.0 &   1.74 \\
       & 5226.543 &   1.57 &  $-$1.23 &   29.2 &   1.68 \\
       & 5336.771 &   1.58 &  $-$1.63 &   14.4 &   1.71 \\
Cr I   & 4254.332 &   0.00 &  $-$0.11 &   69.3 &   1.37 \\
       & 4274.796 &   0.00 &  $-$0.23 &   58.4 &   1.31 \\
       & 4289.716 &   0.00 &  $-$0.36 &   56.0 &   1.40 \\
       & 5206.038 &   0.94 &   0.02 &   37.2 &   1.76 \\
       & 5208.419 &   0.94 &   0.16 &   43.1 &   1.72 \\
       & 5409.772 &   1.03 &  $-$0.72 &    9.5 &   1.87 \\
Mn I   & 4030.753 &   0.00 &  $-$0.48 &   60.2 &   1.17 \\
       & 4033.062 &   0.00 &  $-$0.62 &   46.3 &   1.15 \\
       & 4034.483 &   0.00 &  $-$0.81 &   35.3 &   1.19 \\
Fe I   & 3763.789 &   0.99 &  $-$0.24 &  129.6 &   3.70 \\
       & 3767.192 &   1.01 &  $-$0.39 &  110.6 &   3.40 \\
       & 3786.677 &   1.01 &  $-$2.23 &   45.0 &   3.80 \\
       & 3787.880 &   1.01 &  $-$0.86 &  107.3 &   3.77 \\
       & 3815.840 &   1.48 &   0.24 &  125.2 &   3.67 \\
       & 3820.425 &   0.86 &   0.12 &  158.0 &   3.70 \\
       & 3824.444 &   0.00 &  $-$1.36 &  150.2 &   4.05 \\
       & 3825.881 &   0.91 &  $-$0.04 &  136.8 &   3.54 \\
       & 3827.823 &   1.56 &   0.06 &  103.2 &   3.38 \\
       & 3840.438 &   0.99 &  $-$0.51 &  120.4 &   3.71 \\
       & 3849.967 &   1.01 &  $-$0.97 &  122.2 &   4.24 \\
       & 3856.372 &   0.05 &  $-$1.29 &  142.2 &   3.87 \\
       & 3859.911 &   0.00 &  $-$0.71 &  167.4 &   3.66 \\
       & 3865.523 &   1.01 &  $-$0.98 &  116.8 &   4.10 \\
       & 3878.018 &   0.96 &  $-$0.91 &  111.6 &   3.83 \\
       & 3886.282 &   0.05 &  $-$1.08 &  152.0 &   3.84 \\
       & 3887.048 &   0.91 &  $-$1.14 &  102.7 &   3.77 \\
       & 3895.656 &   0.11 &  $-$1.67 &  132.3 &   4.09 \\
       & 3899.707 &   0.09 &  $-$1.53 &  137.2 &   4.04 \\
       & 3920.258 &   0.12 &  $-$1.75 &  129.2 &   4.10 \\
       & 3922.912 &   0.05 &  $-$1.65 &  131.8 &   3.97 \\
       & 4005.242 &   1.56 &  $-$0.61 &   94.4 &   3.77 \\
       & 4045.812 &   1.48 &   0.28 &  133.8 &   3.72 \\
       & 4063.594 &   1.56 &   0.07 &  110.2 &   3.46 \\
       & 4071.738 &   1.61 &  $-$0.02 &  118.4 &   3.81 \\
       & 4132.058 &   1.61 &  $-$0.67 &   79.4 &   3.52 \\
       & 4143.868 &   1.56 &  $-$0.46 &  105.4 &   3.84 \\
       & 4187.039 &   2.45 &  $-$0.55 &   34.5 &   3.61 \\
       & 4187.795 &   2.42 &  $-$0.55 &   45.2 &   3.76 \\
       & 4191.431 &   2.47 &  $-$0.73 &   35.5 &   3.83 \\
       & 4199.095 &   3.05 &   0.25 &   31.5 &   3.46 \\
       & 4202.029 &   1.48 &  $-$0.70 &   84.5 &   3.48 \\
       & 4222.213 &   2.45 &  $-$0.97 &   26.7 &   3.87 \\
       & 4233.603 &   2.48 &  $-$0.60 &   25.2 &   3.50 \\
       & 4250.119 &   2.47 &  $-$0.40 &   48.1 &   3.71 \\
       & 4260.474 &   2.40 &  $-$0.02 &   68.8 &   3.60 \\
       & 4271.154 &   2.45 &  $-$0.35 &   45.7 &   3.59 \\
       & 4271.761 &   1.48 &  $-$0.16 &  107.9 &   3.46 \\
       & 4282.403 &   2.17 &  $-$0.82 &   36.4 &   3.57 \\
       & 4337.046 &   1.56 &  $-$1.70 &   52.0 &   3.98 \\
       & 4383.545 &   1.48 &   0.20 &  125.7 &   3.49 \\
       & 4404.750 &   1.56 &  $-$0.14 &  113.2 &   3.62 \\
       & 4415.123 &   1.61 &  $-$0.61 &  101.2 &   3.87 \\
       & 4447.717 &   2.22 &  $-$1.34 &   27.2 &   3.96 \\
       & 4461.653 &   0.09 &  $-$3.20 &   85.4 &   4.26 \\
       & 4494.563 &   2.20 &  $-$1.14 &   31.9 &   3.83 \\
       & 4871.318 &   2.87 &  $-$0.36 &   27.0 &   3.71 \\
       & 4872.138 &   2.88 &  $-$0.57 &   21.1 &   3.80 \\
       & 4891.492 &   2.85 &  $-$0.11 &   33.8 &   3.57 \\
       & 4918.994 &   2.87 &  $-$0.34 &   21.1 &   3.56 \\
       & 4920.503 &   2.83 &   0.07 &   45.9 &   3.57 \\
       & 4939.687 &   0.86 &  $-$3.34 &   10.6 &   3.81 \\
       & 4994.130 &   0.92 &  $-$3.08 &   19.5 &   3.92 \\
       & 5041.072 &   0.96 &  $-$3.09 &   27.5 &   4.16 \\
       & 5049.820 &   2.28 &  $-$1.36 &   16.9 &   3.75 \\
       & 5083.339 &   0.96 &  $-$2.96 &   21.4 &   3.89 \\
       & 5123.720 &   1.01 &  $-$3.07 &   21.8 &   4.07 \\
       & 5150.840 &   0.99 &  $-$3.04 &   16.8 &   3.88 \\
       & 5151.911 &   1.01 &  $-$3.32 &   12.5 &   4.03 \\
       & 5166.282 &   0.00 &  $-$4.20 &   27.9 &   4.10 \\
       & 5171.596 &   1.49 &  $-$1.79 &   49.2 &   3.86 \\
       & 5192.344 &   3.00 &  $-$0.42 &   16.3 &   3.64 \\
       & 5194.942 &   1.56 &  $-$2.09 &   33.7 &   3.99 \\
       & 5216.274 &   1.61 &  $-$2.15 &   21.5 &   3.85 \\
       & 5232.940 &   2.94 &  $-$0.06 &   29.1 &   3.52 \\
       & 5266.555 &   3.00 &  $-$0.39 &   16.8 &   3.62 \\
       & 5269.537 &   0.86 &  $-$1.32 &  121.8 &   3.94 \\
       & 5324.179 &   3.21 &  $-$0.24 &   16.8 &   3.72 \\
       & 5328.039 &   0.92 &  $-$1.47 &  108.7 &   3.86 \\
       & 5328.532 &   1.56 &  $-$1.85 &   32.3 &   3.72 \\
       & 5371.490 &   0.96 &  $-$1.65 &  107.6 &   4.05 \\
       & 5397.128 &   0.92 &  $-$1.99 &   85.2 &   3.92 \\
       & 5405.775 &   0.99 &  $-$1.84 &   88.4 &   3.91 \\
       & 5429.697 &   0.96 &  $-$1.88 &   88.8 &   3.91 \\
       & 5434.524 &   1.01 &  $-$2.12 &   71.2 &   3.92 \\
       & 5446.917 &   0.99 &  $-$1.91 &   89.5 &   3.99 \\
       & 5455.609 &   1.01 &  $-$2.09 &   71.6 &   3.90 \\
       & 5497.516 &   1.01 &  $-$2.85 &   24.2 &   3.88 \\
       & 5506.779 &   0.99 &  $-$2.80 &   32.7 &   3.97 \\
       & 5586.756 &   3.37 &  $-$0.14 &   13.1 &   3.66 \\
       & 6393.601 &   2.43 &  $-$1.58 &   11.2 &   3.87 \\
Fe II  & 4233.172 &   2.58 &  $-$1.90 &   35.3 &   3.66 \\
       & 4923.927 &   2.89 &  $-$1.50\tablenotemark{a} &   47.7 &   3.79 \\
Co I   & 3845.461 &   0.92 &   0.01 &   59.5 &   1.50 \\
       & 3995.302 &   0.92 &  $-$0.22 &   50.5 &   1.53 \\
       & 4118.767 &   1.05 &  $-$0.49 &   23.1 &   1.52 \\
       & 4121.311 &   0.92 &  $-$0.32 &   40.8 &   1.51 \\
Ni I   & 3775.565 &   0.42 &  $-$1.39\tablenotemark{a} &   68.2 &   2.37 \\
       & 3807.138 &   0.42 &  $-$1.18 &   75.9 &   2.31 \\
       & 3858.292 &   0.42 &  $-$0.97 &   93.8 &   2.51 \\
       & 5476.900 &   1.83 &  $-$0.89 &   15.5 &   2.38 \\
Sr II  & 4077.710 &   0.00 &   0.16 &   70.0 &  $-$2.15 \\
       & 4215.520 &   0.00 &  $-$0.16 &   51.5 &  $-$2.17 \\
Ba II  & 4934.076 &   0.00 &  $-$0.15 &   40.5 &  $-$2.20 \\
       & 6141.730 &   0.70 &  $-$0.08 &   14.4 &  $-$2.09 \\
       & 6496.910 &   0.60 &  $-$0.38 &   13.6 &  $-$1.96 \\
\enddata
\tablenotetext{a}{$\log gf$ from VALD}
\end{deluxetable}

\clearpage

\begin{deluxetable}{lrrrr} 
\tablecolumns{5} 
\tablewidth{0pc} 
\tablecaption{1D LTE ABUNDANCES OF BOO--1137} 
\tablehead{ 

\colhead{Species} & 
\colhead{N$_{\mbox{\scriptsize lines}}$} & 
\colhead{$\log\epsilon (\mbox{X})$} & 
\colhead{s.e.$_{\log\epsilon}$\tablenotemark{a}} & 

\colhead{[X/Fe]} \\
\colhead{(1)} & \colhead{(2)} & \colhead{(3)} & \colhead{(4)} & \colhead{(5)}
}
\startdata 
  C(CH)       &  syn\tablenotemark{b} & $   4.98 $ & $   0.20 $ & $   0.25 $ \\
  N(NH)       &  syn\tablenotemark{b} & $   5.22 $ & $   0.30 $ & $   1.10 $ \\
  \ion{O}{1}  &    1 & $ < 6.94 $ & $    ... $ & $ < 1.94 $ \\
  \ion{Na}{1} &    2 & $   2.45 $ & $   0.05 $ & $  -0.06 $ \\
  \ion{Mg}{1} &    8 & $   4.34 $ & $   0.10 $ & $   0.47 $ \\
  \ion{Al}{1} &    2 & $   2.10 $ & $   0.11 $ & $  -0.61 $ \\
  \ion{Si}{1} &    1 & $   4.62 $ & $    ... $ & $   0.77 $ \\
  \ion{Ca}{1} &    9 & $   3.17 $ & $   0.08 $ & $   0.52 $ \\
  \ion{Sc}{2} &    3 & $  -0.77 $ & $   0.12 $ & $  -0.16 $ \\
  \ion{Ti}{1} &    5 & $   1.84 $ & $   0.04 $ & $   0.60 $ \\
  \ion{Ti}{2} &   19 & $   1.76 $ & $   0.05 $ & $   0.52 $ \\
  \ion{Cr}{1} &    6 & $   1.57 $ & $   0.10 $ & $  -0.41 $ \\
  \ion{Mn}{1} &    3 & $   1.17 $ & $   0.02 $ & $  -0.56 $ \\
  \ion{Fe}{1} &   81 & $   3.79 $ & $   0.02 $ & $  -3.66\tablenotemark{c}  $ \\
  \ion{Fe}{2} &    2 & $   3.73 $ & $   0.08 $ & $  -0.06 $ \\
  \ion{Co}{1} &    4 & $   1.51 $ & $   0.02 $ & $   0.25 $ \\
  \ion{Ni}{1} &    4 & $   2.39 $ & $   0.05 $ & $  -0.18 $ \\
  \ion{Zn}{1} &    1 & $ < 1.64 $ & $    ... $ & $ < 0.70 $ \\
  \ion{Sr}{2} &    2 & $  -2.16 $ & $   0.02 $ & $  -1.42 $ \\
  \ion{Ba}{2} &    3 & $  -2.08 $ & $   0.08 $ & $  -0.59 $ \\
  \ion{Eu}{2} &    1 & $ <-2.65 $ & $    ... $ & $ < 0.49 $ \\
\enddata 
\tablenotetext{a}{Uncertainty of the fit in the case of spectrum synthesis; standard
  error of the mean for species having at least two line strength measurements}
\tablenotetext{b}{Determined using spectrum synthesis}
\tablenotetext{c}{The tabulated value is [Fe/H]}
\end{deluxetable} 

\clearpage                                                                                                                                                                
\begin{deluxetable}{lrrrrr}                                                                                                                                               
\tablecolumns{6}                                                                                                                                                          
\tablewidth{0pt}                                                                                                                                                          
\tablecaption{ABUNDANCE ERRORS FROM UNCERTAINTIES IN ATMOSPHERIC PARAMETERS}                                                                                              
\tablehead{                                                                                                                                                               
  \colhead{Species}    & \colhead{$\Delta${\teff}}   & \colhead{$\Delta${\logg}}   & \colhead{$\Delta$[M/H]}   & \colhead{$\Delta\xi_t$} & \colhead{$\Delta$[X/Fe]} \\   
  \colhead{}           & \colhead{(200K)}  & \colhead{(0.3~dex)}  & \colhead{(0.3~dex)} & \colhead{(0.2~{\kms})} & \colhead{(dex )}   \\                                  
  \colhead{(1)}        & \colhead{(2)}    & \colhead{(3)}    & \colhead{(4)}   &\colhead{(5)}    & \colhead{(6)}                                                          
  }                                                                                                                                                                       
\startdata                                                                                                                                                                
C              & $    0.28  $ & $       -0.08  $ & $        0.00  $ & $        0.00  $ & $        0.29 $ \\   
N              & $    0.38  $ & $       -0.08  $ & $        0.00  $ & $        0.00  $ & $        0.39 $ \\   
\ion{Na}{1}    & $   -0.02  $ & $        0.00  $ & $        0.00  $ & $       -0.03  $ & $        0.03 $ \\   
\ion{Mg}{1}    & $   -0.10  $ & $        0.01  $ & $        0.00  $ & $        0.00  $ & $        0.10 $ \\   
\ion{Al}{1}    & $    0.02  $ & $       -0.01  $ & $        0.00  $ & $       -0.06  $ & $        0.06 $ \\   
\ion{Si}{1}    & $    0.00  $ & $       -0.07  $ & $       -0.02  $ & $       -0.14  $ & $        0.16 $ \\   
\ion{Ca}{1}    & $   -0.10  $ & $        0.00  $ & $        0.00  $ & $       -0.01  $ & $        0.10 $ \\   
\ion{Sc}{2}    & $   -0.08  $ & $        0.10  $ & $        0.01  $ & $       -0.01  $ & $        0.13 $ \\   
\ion{Ti}{1}    & $    0.04  $ & $       -0.01  $ & $       -0.01  $ & $       -0.01  $ & $        0.04 $ \\   
\ion{Ti}{2}    & $   -0.12  $ & $        0.11  $ & $        0.01  $ & $        0.00  $ & $        0.16 $ \\   
\ion{Cr}{1}    & $    0.02  $ & $        0.00  $ & $        0.00  $ & $        0.00  $ & $        0.02 $ \\   
\ion{Mn}{1}    & $    0.08  $ & $       -0.01  $ & $       -0.01  $ & $       -0.01  $ & $        0.08 $ \\   
\ion{Fe}{1}    & $    0.11\tablenotemark{a}  $ & $       -0.02\tablenotemark{a}  $ & $        0.00\tablenotemark{a}  $ & $        0.00\tablenotemark{a}  $ & $        0.11\tablenotemark{a} $ \\   
\ion{Fe}{2}    & $   -0.20  $ & $        0.10  $ & $        0.00  $ & $       -0.02  $ & $        0.22 $ \\   
\ion{Co}{1}    & $    0.06  $ & $        0.00  $ & $        0.00  $ & $       -0.01  $ & $        0.06 $ \\   
\ion{Ni}{1}    & $    0.02  $ & $        0.00  $ & $        0.00  $ & $        0.00  $ & $        0.02 $ \\   
\ion{Sr}{2}    & $   -0.06  $ & $        0.10  $ & $        0.01  $ & $       -0.03  $ & $        0.12 $ \\   
\ion{Ba}{2}    & $   -0.06  $ & $        0.10  $ & $        0.01  $ & $       -0.01  $ & $        0.12 $ \\ 
\enddata                                                                                                      
\tablenotetext{a}{Errors pertain to uncertainties in [Fe/H]}
\end{deluxetable}                                                                                             

\begin{deluxetable}{lccccccc}
\tablecolumns{8}
\tablewidth{0pt}
\tablecaption{MONTE-CARLO ANALYSIS OF BOO-1137 $\alpha$-ELEMENT RELATIVE ABUNDANCES}
\tablehead{

\colhead{Species}    & \colhead{RMS$_{\rm Halo}$\tablenotemark{a}}    & \colhead{[X/Fe]$_{\rm Halo}$\tablenotemark{b}}     & \colhead{[X/Fe]$_{\rm Boo}$}     & \colhead{$\sigma$[X/Fe]$_{\rm Boo}$}     & \colhead{$\Delta$[X/Fe]}    & \colhead{$\Delta$[X/Fe]/}                   & \colhead{Fraction}  \\  
\colhead{}           & \colhead{}                    & \colhead{}                        & \colhead{}                            & \colhead{}                                    & \colhead{}                  & \colhead{$\sigma$[X/Fe]$_{\rm Boo}$}    & \colhead{}   \\                                  
\colhead{(1)}        & \colhead{(2)}     & \colhead{(3)}                   & \colhead{(4)}                      &   \colhead{(5)}                           & \colhead{(6)}            & \colhead{(7)} & \colhead{(8)}  }
\startdata
\ion{Mg}{1}          & 0.11    & 0.250   &  0.470    & 0.14    & 0.220   & 1.57  & 0.108 \\
\ion{Si}{1}          & 0.20    & 0.412   &  0.770    & 0.26    & 0.358   & 1.38  & 0.138 \\
\ion{Ca}{1}          & 0.11    & 0.290   &  0.520    & 0.13    & 0.230   & 1.77  & 0.089 \\
\ion{Ti}{1}          & 0.09    & 0.346   &  0.600    & 0.10    & 0.254   & 2.54  & 0.029 \\ 
\ion{Ti}{2}          & 0.11    & 0.234   &  0.520    & 0.17    & 0.286   & 1.68  & 0.079 \\ 
$\alpha$             & 0.09    & 0.274   &  0.516    & 0.07    & 0.242   & 3.40  & 0.017 \\

\enddata                                                                                                      
\tablenotetext{a}{From Cayrel et al.\ (2004) and the present work}

\tablenotetext{b}{Halo value at [Fe/H] = --3.66, determined from the regressions lines of
Cayrel et al.\ (2004) and the present work.}

\end{deluxetable}                                                                                  

\clearpage

\clearpage
\begin{figure}[htbp]
\vspace{1cm}
\begin{center}
\includegraphics[width=6.0cm,angle=0]{f2.eps}

  \caption{\label{Fig:Eqwid} (a,b) Comparison of equivalent widths
  measured by D.Y. and J.E.N. in the VLT/UVES spectrum of {\boos}.
  ((b) contains the linear least squares best fit and 1--1 lines, while
  in (a) $\Delta$W$_{\lambda}$ = W$_{\lambda}$(J.E.N.) --
  W$_{\lambda}$(D.Y.).) (c) The equivalent widths of {\cd} ([Fe/H] =
  --4.2) and {\bd} ([Fe/H] = --3.1) (from Cayrel et al.\ 2004) versus
  the present results for {\boos}.  Note that the line strengths in
  {\cd} and {\bd} are smaller and larger, respectively, than in
  {\boos}.}

\end{center}
\end{figure}

\clearpage
\begin{figure}[htbp]
\vspace{1cm}
\begin{center}
\includegraphics[width=15cm,angle=0]{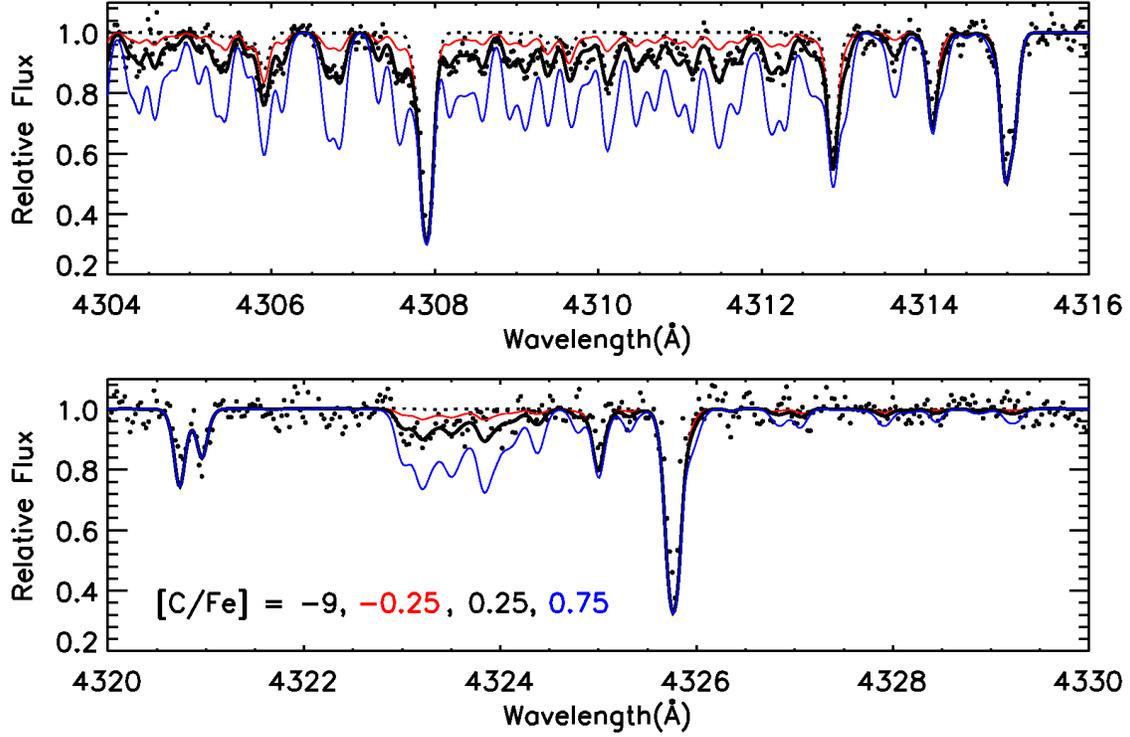}

  \caption {\label{Fig:Carbon} Comparison of the spectrum of {\boos}
  (heavy black dots) in the regions 4304--4316~{\AA} (upper panel) and
  4320--4330~{\AA} (lower panel) with synthetic spectra including CH
  A--X lines for relative carbon abundances [C/Fe] = --9.0 (thin
  dotted line), --0.25 (thin red line), 0.25 (thick line; best fit), and
  0.75 (thin blue line)}

\end{center}
\end{figure}

\clearpage
\begin{figure}[htbp]
\vspace{1cm}
\begin{center}
\includegraphics[width=15cm,angle=0]{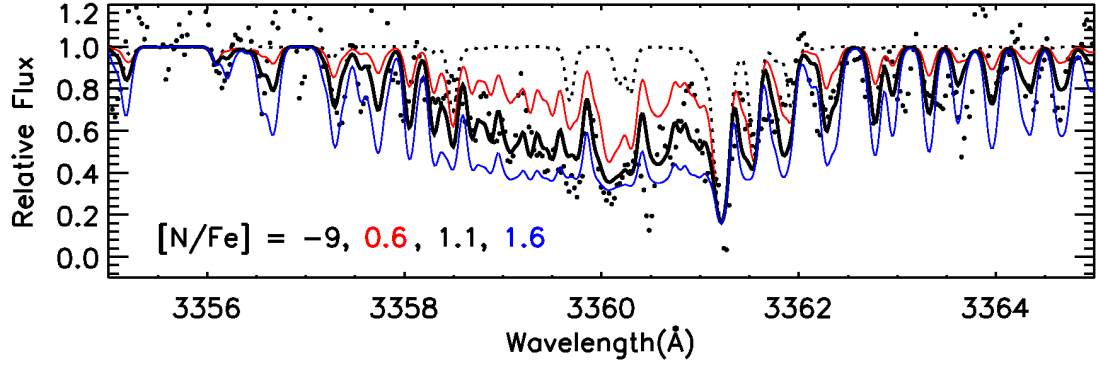}

  \caption {\label{Fig:Nitrogen} The spectrum of {\boos} (heavy black
  dots) in the region of NH A--X band at 3360~{\AA}, compared with
  synthetic spectra having relative abundances [N/Fe] = --9.0 (thin
  dotted line), 0.6 (thin red line), 1.1 (thick line; best fit), and
  1.6 (thin blue line).}

\end{center}
\end{figure}

\clearpage
\begin{figure}[htbp]
\vspace{1cm}
\begin{center}
\includegraphics[width=15cm,angle=0]{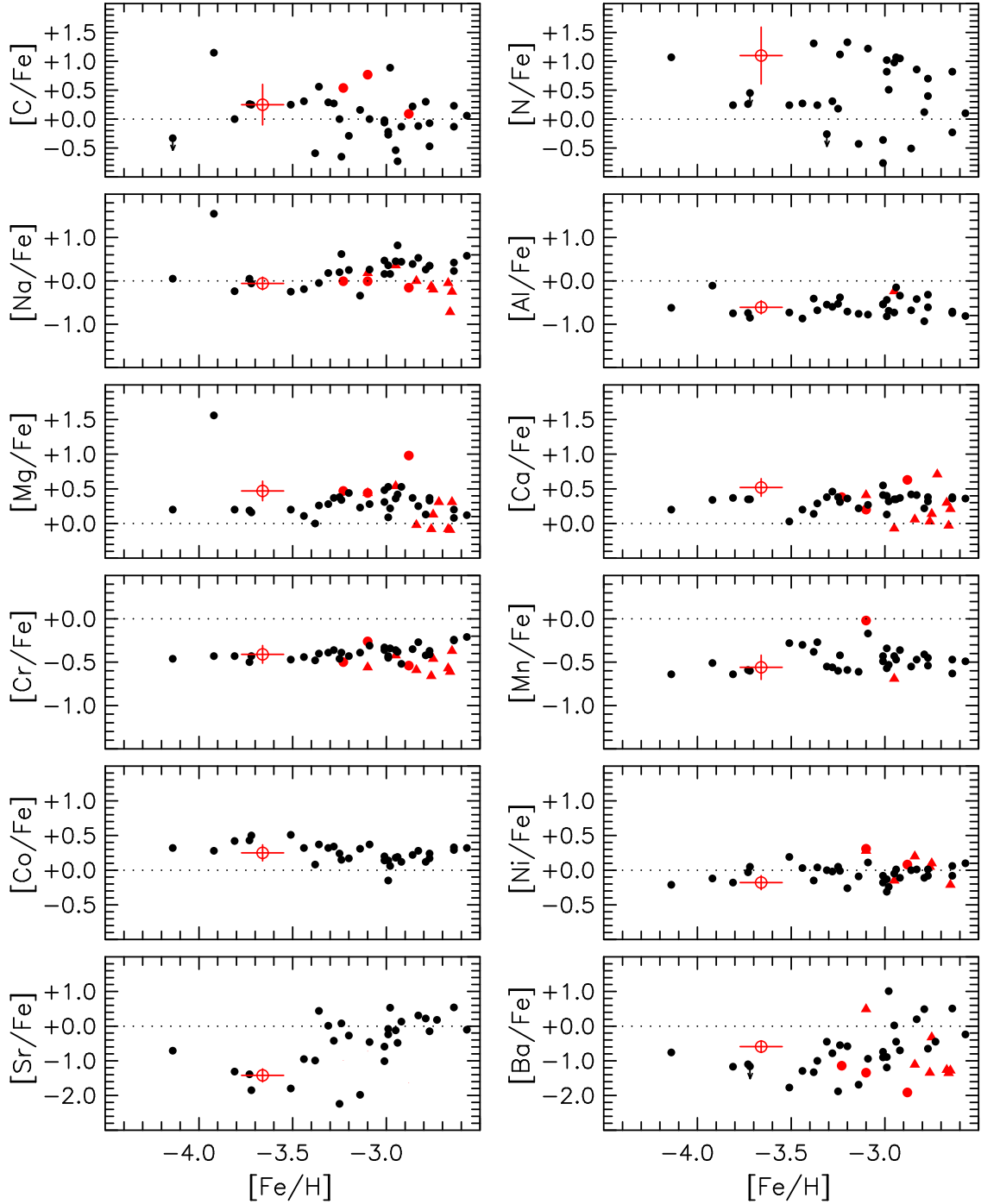}

\caption {\label{Fig:XFe} Relative abundances, [X/Fe], as a function
of [Fe/H].  {\boos} is represented by an open red circle, while filled
red circles stand for giants in the ultra-faint dSph Com Ber and U~Ma~II
(Frebel et al.\ 2009) and the filled red triangles for stars in
the more luminous Sextans dSph (Aoki et al.\ 2009).  The
filled black circles are from Cayrel et al.\ (2004) (Na--Ni), Spite et
al.\ (2005) (C and N), and Fran{\c c}ois et al.\ (2007) (Sr and Ba),
respectively.}

\end{center}
\end{figure}

\clearpage
\begin{figure}[htbp]
\vspace{1cm}
\begin{center}
\includegraphics[width=7.0cm,angle=0]{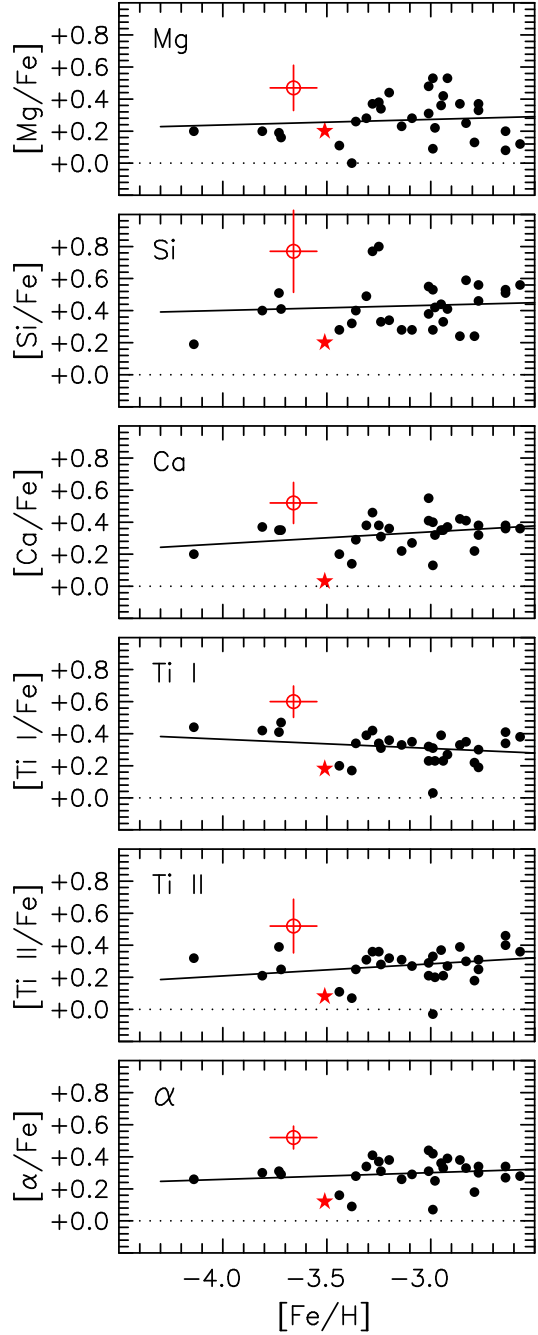}

  \caption {\label{Fig:Alphas} Relative abundances [X/Fe] for the
  $\alpha$-elements as a function of [Fe/H], where we use
  [$\alpha$/Fe] to denote the average of the relative abundances of
  [Mg/Fe], [Ca/Fe], [Ti I/Fe], and [Ti II/Fe].  {\boos} is represented
  by an open red circle, while the filled black circles and the red
  star (for CS22968--014) are from Cayrel et al. (2004).  See text for
  discussion.}

\end{center}
\end{figure}

\end{document}